\documentclass[aps,prl,preprint,groupedaddress,preprintnumbers]{revtex4-1}
\usepackage{graphicx}

\begin{document}
\preprint{CENPA 16-v2}
%\preprint{}

\title{Neutrino Oscillations and Entanglement}
%v2.5 

\author{R.\,G.\,Hamish Robertson}
\email[]{rghr@uw.edu}
%\thanks{Research supported by DOE under grant DE-FG02-97ER41020}
\affiliation{Department of Physics \\
and Center for Experimental Nuclear Physics and Astrophysics, \\
 University of Washington,  Seattle, WA 98195 \\}

\date{\today}

\begin{abstract}

   An improved treatment of neutrino oscillations follows when neutrino mass eigenstates and their associated recoiling particle states are entangled.  When the neutrino and its recoil partner are treated as a single entangled quantum state the conservation of energy and momentum in neutrino oscillations can be assured, even in a plane-wave treatment.   The oscillation wavelength between the neutrino and its associated recoil partner emerges as the fundamental periodicity in this analysis.  However, the experimentally determined oscillation wavelength for neutrinos detected at some distance from a known source region is still in all cases given by the standard expression in current use.   

\end{abstract}

%\keywords{}

\maketitle
\section{Introduction}

Neutrinos produced in a weak charged-current decay are prepared in a state of specified flavor because they are accompanied by a charged lepton with that  flavor.  As is now well established, such flavor states are in fact a linear superposition of mass eigenstates of mixed flavor.  The mass differences lead to a phase slip between the quantum waves of each mass component, which can be experimentally detected as a change in flavor at some place spatially separated from the point of decay.

Neutrino oscillations, a phenomenon recognized in Pontecorvo's farsighted prediction in 1957 \cite{Pontecorvo:1957cp}, can be treated in a simple and straightforward way leading to a correct expression for the wavelength in terms of the neutrino energy and the difference between the squares of the masses of the eigenstates.  But a propagating neutrino that consists of 2 or more mass eigenstates cannot simultaneously have a well defined energy and a well defined momentum.  How  Lorentz transformations should be made then also depends on which is chosen.   These difficulties are well known -- a thorough summary has been given recently by Akhmedov and Smirnov \cite{Akhmedov:2009rb}.  Generally the solution is to introduce a wave-packet approach, which is necessary for other reasons in any case, and treat both the energy and the momentum as somewhat undefined.  Another strategy is to use quantum field theory, treating the neutrino as a propagator and thereby avoiding detailed consideration of an internal line, but also rendering it difficult to obtain oscillations at all \cite{Keister:2009qn,Akhmedov:2010ms}.

Such difficulties do not arise when the neutrino and its recoil are treated in a unified way as entangled-state components.  The role of entanglement in neutrino and neutral kaon oscillations has been considered by others \cite{Cohen:2008qb,Dolgov:1997xr,Burkhardt:2003cz,Burkhardt:1998zj,Lowe:1996ct,Wu:2010yr,Keister:2009qn,Goldman:2010zz,Wu:2010tr}, who, with some exceptions \cite{Wu:2010yr,Wu:2010tr}, show that the standard formulae are obtained.  In our analysis, an oscillation wavelength between the neutrino and its associated recoiling partner emerges.  When a source decay region is then localized, that oscillation wavelength can be related to a laboratory oscillation wavelength for the neutrino distance from the source, which can be shown \cite{Vogel,Kayser} to be the same as the standard wavelength.

The basic principles are best seen in two-body decays, such as pion decay, or electron capture.    Two-body decays are interesting for another reason: --  the kinematics that produce neutrino flavor oscillations produce, under restrictive conditions (of which one  is that the neutrino be observed), an `oscillation' of the recoil particle or nucleus.

\section{Calculation}\label{S:cal}

Consider a two-body
decay at rest, for example 
$^{7}$Be + e$^-$
$\rightarrow \  ^{7}$Li + $\nu_e$. 
The final state is a two-particle state comprising the neutrino and the recoiling $^{7}$Li ion:     \begin{eqnarray}
\left |R{\nu_e}\right \rangle \sim  \left |R\right \rangle\left |{\nu_e}\right \rangle\rho^{-1}\exp{i(p\rho-Et)}.\label{eqone}
\end{eqnarray}
It has the form of a product wave function of the internal degrees of freedom of the neutrino and recoil particle and a function of their relative motion, a spherical wave.
The energy $E$ is the energy of the final state, which is the same as the energy of the initial state.    The momentum of each particle is $p$ and their separation is $\rho$.  This is the basic form of the two-body wave function that will be recast in Lorentz invariant form below.  The neutrino and recoil particle are entangled in a single quantum state.  A key feature of this entanglement is the appearance of a single conserved energy $E$ and a single time coordinate $t$.  

The (electron) neutrino is a linear combination of 
mass eigenstates
$\nu_i$ with masses
$m_i$,
\begin{eqnarray}
\left | \nu_e \right \rangle = \sum_i U_{ei}^*\left | \nu_i \right \rangle,
\end{eqnarray}
where the $U_{ei}$ are elements of the Maki-Nakagawa-Sakata-Pontecorvo mixing matrix \cite{Amsler:2008zz}.  In the rest frame of $^{7}$Be (mass  $M$), decaying into an electron neutrino $\nu_e$
and a recoiling particle (henceforth simply `the recoil', in this case $^{7}$Li)  with mass  $m$,  the neutrino (recoil) energies are \cite{Amsler:2008zz}
\begin{eqnarray}
E_{i(Ri)} = \frac{ M^2 -(+) m^2 +(-) m_i^2}{2M}. 
\end{eqnarray}
%and the recoil energies are, correspondingly, 
%\begin{eqnarray}
%E_{Ri} = \frac{ M^2 + m^2 - m_i^2}{2M}.  \label{eqthree}
%\end{eqnarray}
 The sum of neutrino and recoil energy is always $M$ in the rest frame (the masses include the appropriate numbers of atomic electrons).   The momenta are equal and opposite for the recoil and neutrino eigenstate,
\begin{eqnarray}
|{\bf p}_i| &=& \frac{ ([M^2 - (m + m_i)^2][M^2 - (m - m_i)^2])^{1/2}}{
2M}. 
\end{eqnarray}
A complete description of the final state  must include the wave functions of both the neutrino, which may have a general flavor projection $\left |{\nu}\right \rangle$, and the recoil.  The wave function  can be written schematically in terms of the four-momentum $P$ and the spacetime coordinates $X$ of the recoil, and the corresponding quantities $p$ and $x$ for the neutrino:
\begin{eqnarray}
\left |R{\nu}\right \rangle &\sim&  \sum_iU_{ei}^*\left |R(P,X)\right \rangle_i\left |{\nu_i}(p,x)\right \rangle_i.
\end{eqnarray}
Kinematics requires the subscript $i$ on the recoil wavefunction.       The recoil and the neutrino move as free particles, asymptotically in plane-wave states.
\begin{eqnarray}
&\left |R{\nu}\right \rangle^{rs} =  \sum_{i}U_{ei}^*\left |R\right \rangle\left |{\nu_i}\right \rangle\times \nonumber\\
&\left[w^r({\bf P})e^{-i\epsilon_rP_\mu X^\mu} w^s({\bf p})e^{-i\epsilon_sp_\mu x^\mu}\right]_i. \label{eqseven}
\end{eqnarray}
The $w$ are spinors, $\epsilon = +1(-1)$ identifies positive-energy (negative-energy) states, and ${\bf P, p}$ are the 3-momenta of the recoil and neutrino respectively \cite{Bjorken:1964zz}.  The particles move along the $x$-axis.  The spin degrees of freedom and negative-energy states are irrelevant for the present, kinematic, considerations.   In the rest frame of the parent, ${\bf P} = -{\bf p}$.  
\begin{eqnarray}
\left |R{\nu}\right \rangle & \sim &  \sum_iU_{ei}^*\left |R\right \rangle\left |{\nu_i}\right \rangle \left[e^{-iE_R t - ip_xX}e^{-iE_\nu t + ip_xx}\right]_i \nonumber \\
&=& \left |R\right \rangle e^{-iE_at}\sum_iU_{ei}^*\left |{\nu_i}\right \rangle \left[e^{ ip_x(-X+x)}\right]_i. 
\end{eqnarray}
where $E_a= M$ is the energy of the parent.  The final state is seen to have the desired form of a product wave function of the internal degrees of freedom of the neutrino and recoil, and a function of their relative motion, as in Eq.~\ref{eqone}.

Viewed from a frame moving at velocity $-\beta$ along the x-axis, the momenta have the transformed values $P_x'=\gamma\beta E_R - \gamma p_x$ and $p_x'=\gamma\beta E_\nu + \gamma p_x$,  the space coordinates become $X'=\gamma\beta t + \gamma X$ and $x'=\gamma\beta t + \gamma x$, and the time coordinates become $t_\nu'=\gamma t + \gamma\beta x$ and $t_R'=\gamma t + \gamma \beta X$.  The energies become $E_R'=\gamma E_R - \gamma\beta p_x$ and $E_\nu'=\gamma E_\nu + \gamma \beta p_x$.  The primed quantities represent the quantities in the laboratory frame when the parent is boosted to a velocity $\beta$.  Since the arguments of the exponentials in Eq.~\ref{eqseven} are Lorentz invariants, one may equivalently write
\begin{eqnarray}
&{\left |R{\nu}\right \rangle^{rs}}'  = \sum_{i}U_{ei}^*\left |R\right \rangle\left |{\nu_i}\right \rangle \times \nonumber\\
&\left[w^r({\bf P}')e^{-i\epsilon_rP_\mu' {X^\mu}'} w^s({\bf p'})e^{-i\epsilon_sp_\mu' {x^\mu}'}\right]_i \nonumber\\
&{\left |R{\nu}\right \rangle}' \simeq\sum_iU_{ei}^*\left |R\right \rangle\left |{\nu_i}\right \rangle \left[e^{-iE_R t_R + iP_xX}e^{-iE_\nu t_\nu + ip_xx}\right]_i' \nonumber
\end{eqnarray}

%The equivalence is verified as follows:
%\begin{eqnarray*}
%\left[-E_Rt_R+P_xX-E_\nu t_\nu+p_xx\right]' & = & -(\gamma E_R - \gamma\beta p_x)( \gamma t + \gamma \beta X) \\
%& & + (\gamma\beta E_R - \gamma p_x)(\gamma\beta t + \gamma X) \\
%& & -  (\gamma E_\nu + \gamma \beta p_x)(\gamma t + \gamma\beta x) \\
%& & + (\gamma\beta E_\nu + \gamma p_x)(\gamma\beta t + \gamma x) \\
%& = & \gamma^2(1-\beta^2) (-E_Rt - p_xX - E_\nu t+p_xx) \\
%& = & -E_Rt - p_xX - E_\nu t+p_xx.
%\end{eqnarray*}

If the experimental situation calls for observing the space and/or time coordinates of the particles by detecting them, then  the laboratory spacetime coordinates are of interest.  The phase   
%\begin{widetext}
\begin{eqnarray*}
&\left[-E_Rt_R+P_xX-E_\nu t_\nu+p_xx\right]'  \\
%& -(\gamma E_R - \gamma\beta p_x)t_R'   + (\gamma\beta E_R - \gamma p_x)X'  \\
%& -  (\gamma E_\nu + \gamma \beta p_x)t_\nu'  + (\gamma\beta E_\nu + \gamma p_x)x' \\
%& =  \gamma E_R(- t_R' +\beta X') + \gamma E_\nu(-t_\nu'  +\beta x')   \\
%& + \gamma p_x(\beta [t_R' - t_\nu'] - X' + x') \\
& =  -(E_R+E_\nu)t   + \gamma^{-1} p_x(-X' + x').  
\end{eqnarray*}
%\end{widetext}

The energy $E_R+E_\nu= E_a$ is the energy of the final state in the rest frame, which is the same as the energy of the initial state, and therefore carries no index $i$.  The internal coordinates of the recoil are independent of $i$ as well.   When  the neutrino is detected, a measurement is made of its relative probability density at a specific spacetime location, assuring that the spacetime coordinates of the point of detection are independent of $i$.  The state detected is:
\begin{eqnarray}
\left |R{\nu}\right \rangle' &= & e^{-i E_at}\left |R\right \rangle \sum_iU_{ei}^*\left |{\nu_i}\right \rangle  e^{i \gamma^{-1} p_{x,i}(-X' + x')}.  \label{eqtena}
\end{eqnarray}
The leading phase factor that depends on the energy is unobservable.    As a result, the interference effects of neutrino oscillations arise {\em solely} from the different {\em momenta} in the components in the final state.  (In this we concur with \cite{Lipkin:1995cb}, but see also \cite{Goldman:2010zz}). Since  $p_{x,i} \ne p_{x,j}$,   phase differences exist between the components at the point of detection.  The electron flavor admixture for two admixed components $i,j$ oscillates with a wavelength $\lambda_{ij}'$ given by 
\begin{eqnarray}
& & \frac{ ([M^2 - (m + m_i)^2][M^2 - (m - m_i)^2])^{1/2}}{2\gamma M} \pm \frac{2\pi}{\lambda_{ij}'} \nonumber \\
& =& \frac{ ([M^2 - (m + m_j)^2][M^2 - (m - m_j)^2])^{1/2}}{2\gamma M} \label{eqten}
\end{eqnarray}
 Dropping terms of order $m_i^4$ and writing $m_i^2-m_j^2=\Delta m^2_{ij}$,
\begin{eqnarray}
\frac{1}{\lambda_{ij}'} & \simeq & \frac{\Delta m^2_{ij}}{4\pi \hbar c}\frac{1}{ \gamma M}\frac{M^2+m^2}{M^2-m^2},  \label{eqthtn}
\end{eqnarray}
where the missing units have been made explicit.   All quantities in Eq.~\ref{eqthtn} except $\gamma$ are Lorentz invariants.  In terms of the `neutrino beam energy' $E_0'$, which may without approximation be defined as the energy that massless neutrinos would have in the laboratory, 
\begin{eqnarray}
\frac{1}{\lambda_{ij}'} & \simeq & \frac{\Delta m^2_{ij}}{4\pi \hbar c}\frac{1+\beta}{E_0'}\frac{M^2+m^2}{2M^2}.  \label{eqfrtn}
\end{eqnarray}
Equations \ref{eqthtn} and \ref{eqfrtn} define a wavelength in  laboratory coordinates for  the separation $-X'+x'$ between the neutrino and the recoil \footnote{Goldman \cite{Goldman:2010zz} shows that even the 4th order in neutrino mass can be absorbed in a redefinition of $E_0'$.}.

In the standard expression \cite{Amsler:2008zz} for neutrino oscillation, the survival probability oscillates with a wavelength $\lambda_{ij}'$  given by 
\begin{eqnarray}
\frac{1}{\lambda_{ij}'} & = & \frac{\Delta m^2_{ij}}{4\pi \hbar c E_0' }, \label{eqtwlve}
\end{eqnarray}
It appears the two results are different, but Eq.~\ref{eqfrtn} defines an oscillation length between neutrino and recoil, whereas Eq.~\ref{eqtwlve} is the oscillation length relative to the decay point.  Experimentally, the latter situation is the usual case and so it will be desirable to recast Eq.~\ref{eqfrtn} into an equation for the wavelength of oscillations in $x'$ alone.  This can be done by taking into account the fact that $X'$ and $x'$ are not independent variables, but are related by kinematics once an origin has been defined.  We may conveniently choose the origin in spacetime as the point at which the decay occurs.  In effect, this defines a phase relationship between the plane waves such that the flavor is electron at the origin.   In doing so, we appear to run afoul of the uncertainty principle because plane waves have precise momenta and indefinite positions.    Replacing the single plane-wave states $i,j$ with distributions of plane-wave states having closely similar momenta (by imposing a small boost distribution on the parent), one creates correlated wave-packets.  These superpositions of plane-wave states  nevertheless retain the general properties of interference identified for the single plane-wave states, provided the width of the momentum distribution is not so large that the interference pattern is made indistinct at the point of detection.

The oscillation probability may be found from Eq.~\ref{eqtena}, noting that $\left| R \right\rangle$ describes the internal  coordinates of $R$ and thus is independent of $i$:  
\begin{eqnarray}
P(\nu_e \rightarrow \nu_\beta)&=& \left|\left \langle R\nu_\beta| R \nu_e \right \rangle \right|^2 \nonumber \\
& =& \left|\sum_iU_{ei}^*U_{\beta i}e^{i \gamma^{-1}p_{x,i}(-X'+x')}\right|^2,
\end{eqnarray}
which may be expanded in the usual way \cite{Amsler:2008zz}.  In the oscillation probability, the $i-j$ interference term depends on $(-X'+x')$ through the phase factor $ \exp{[i \gamma^{-1}( p_{x,i}-p_{x,j})(-X'+x')]}$.   It is noteworthy that in the entangled-state picture, oscillations arise even in the plane-wave basis because states containing different neutrino eigenmasses nevertheless have the same {\em total} momentum (zero in the rest frame) and the same total energy. 
 
To find the oscillation probability in the more customary coordinate $x'$, when the decay occurred at or near the origin, the variable $X'$ must be expressed through kinematics in terms of $x'$.  In the rest frame of the parent, the 3-momenta are equal and opposite, and the distances are proportional to the speeds:
\begin{eqnarray}
-X+x & \simeq & x\frac{2M^2}{M^2+m^2}
\end{eqnarray}
where the neutrino masses have been neglected (their contribution turns out to be higher order than $m_i^2$ \cite{Goldman:2010zz,Kayser:2010zz}).  Referring to Eq.~\ref{eqfrtn}, one sees that for $\beta=0$, this replacement yields a wavelength in $x$ that is identical to the standard one, as Vogel has shown \cite{Vogel}.

This equivalence extends, in fact, to all frames, as shown by Kayser \cite{Kayser}.  A general argument will be presented elsewhere \cite{Kayser:2010zz}, but there is an interesting special case for which it is clear.  If the parent is boosted by a velocity that leaves the recoil {\em at rest} in the laboratory, then the wavelength for neutrino oscillations relative to the decay point is given directly by Eq.~\ref{eqfrtn}.  The speed of the recoil in the rest frame (again neglecting the neutrino masses) $|\beta_R|$ is given by
\begin{eqnarray}
1+|\beta_R| & = & \frac{2M^2}{M^2+m^2},
\end{eqnarray} 
and it is evident that for this particular boost, Eq.~\ref{eqfrtn} once again gives a result equivalent to the standard result, Eq.~\ref{eqtwlve}.  That the standard result arises from an entangled-state ansatz, without disentanglement, has been shown also by \cite{Burkhardt:2003cz,Burkhardt:1998zj,Lowe:1996ct}. 

\begin{figure}
 \includegraphics[scale=0.5]{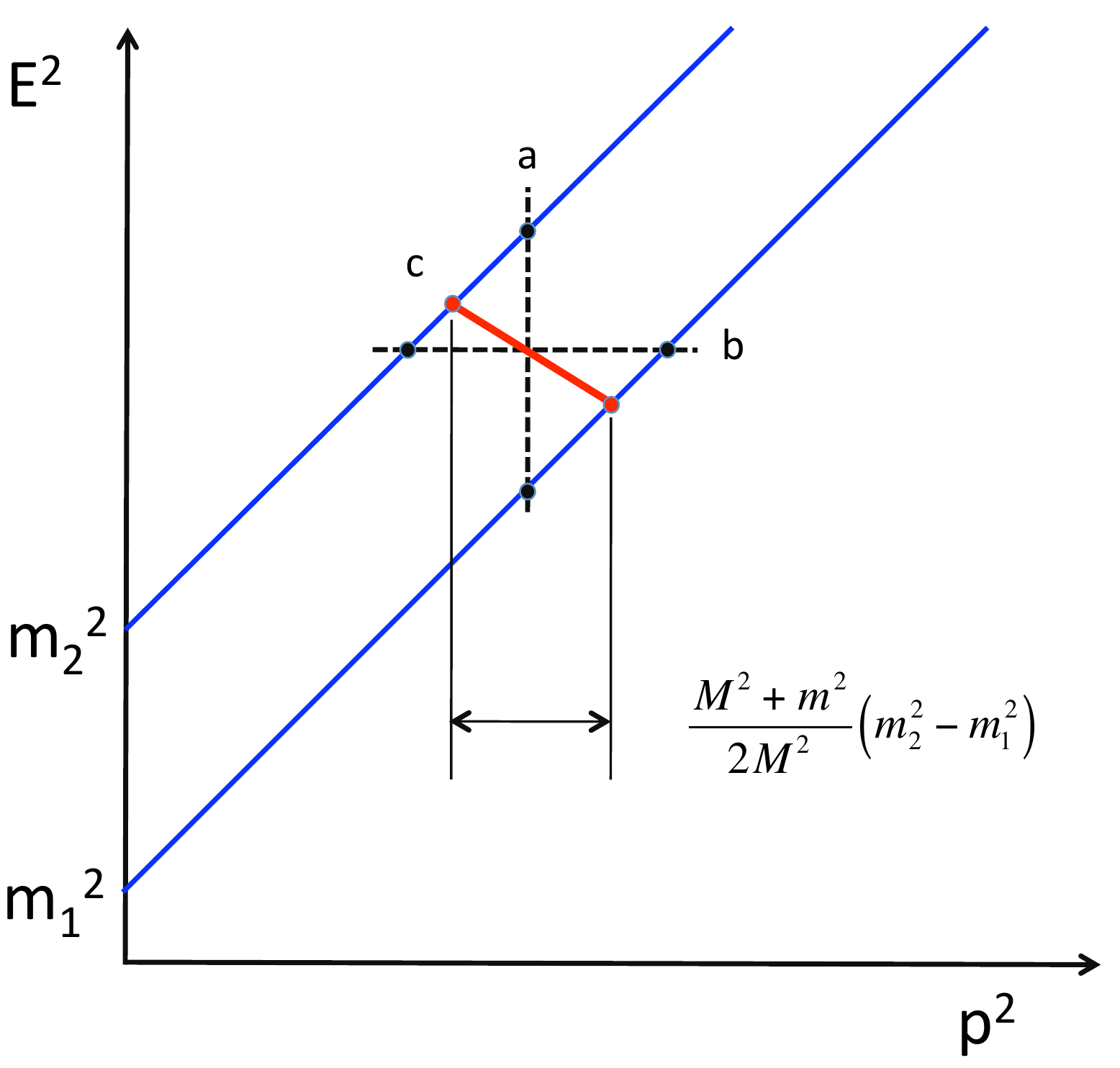}
 \caption{\label{fig:oscfig}Illustration of the relationship between energy and momentum for two neutrino components of mass $m_1$ and $m_2$.  The two lines with slope +1 are the Einstein equation for particles with those masses.  The standard formula for neutrino oscillations may be obtained by selecting a common average momentum (line $a$) or energy (line $b$) and expanding to derive an oscillation phase corresponding to the energy or momentum difference, in time or distance coordinates.  Both happen to give the same result, because the slope is +1 (to second order in .  In the entangled-state picture, the momentum difference between  two neutrino mass eigenstates is given by the difference in 3-momenta at the intersections with line $c$.  The expression displayed shows the modification of $\Delta m^2$ when the phase in the neutrino-recoil system is calculated (in a frame other than the rest frame of the parent, an additional factor  $1+\beta$ enters -- see Eq.~\ref{eqfrtn}).   Despite this constraint, as Cohen {\em et al.}~\cite{Cohen:2008qb} show, the phase for neutrino oscillations relative to the origin is always given by the standard expression, for any momentum/energy combinations allowed by the eigenmasses.}
 \end{figure}

  The entanglement with the recoil enforces a specific relationship between the momenta of the oscillating neutrino components, and does not permit an arbitrary choice (usually either  `beam energy' or `beam momentum') about which to expand in order to derive the oscillation phase.  The total energy, rather than neutrino eigenmass energy, contributes to the state phase, removing it from a physical role. An illustration of the relationship is shown in Fig.~\ref{fig:oscfig}.  

The group velocity of the mixed state in this formalism is equal to the classical particle velocity:
\begin{eqnarray*}
\frac{\Delta E}{\Delta|{\bf p}|} &= & \frac{|{\bf p}|}{E}.
\end{eqnarray*}
The importance of this condition for a physically reasonable description has been emphasized (e.g.   Levy \cite{Levy:2009uz}); here it is derived.

%\section{Recoil Oscillation}

The form of Eq.~\ref{eqtena} shows that oscillation effects are manifested in the separation between the neutrino and the recoil.   Whether `recoil oscillations' are observed depends on whether and how the associated neutrino is detected.  In a realistic situation where the location of the decay is known to some accuracy, the recoil will always be found at a distance from a detected neutrino given by kinematics and it will show no oscillations whether the neutrino is detected in coincidence with the recoil or not.   In a less practical case where the decay point is unknown, the recoil-neutrino distance will show periodic oscillations in space (in the rest frame of the parent).  Our conclusions match those of Dolgov {\em et al.}~\cite{Dolgov:1997xr}.  Consider a toy model of 2-neutrino mixing with equal amplitudes of electron and mu flavor in each eigenmass, i.e.,
\begin{eqnarray}
\left |R{\nu}\right \rangle &\sim & \left |R\right \rangle \left[\left |{\nu_1}\right \rangle +  e^{i \gamma^{-1} (p_{x,2}-p_{x,1})(-X' + x')}\left |{\nu_2}\right \rangle\right]  \label{eqsxtn}
\end{eqnarray}
omitting leading phase factors and constants.  The neutrino may be detected as an electron neutrino, for which the phase is 0, $2\pi$, etc.; as a muon neutrino, for which the phase is $\pi$, $3\pi$, etc.; by a neutral-current process, for which the phase is irrelevant; or not detected at all, for which the phase is again irrelevant.  Detecting the neutrino by a charged-current process enforces a certain phase relationship, and if the recoil is detected in coincidence with the neutrino, its probability will show spatial oscillations as well because the separation $-X' + x'$ between recoil and neutrino may be increased by an integer number of wavelengths to recover the same state.   On the other hand, when the neutrino is detected via the neutral current, or is not detected, then the phase is irrelevant, detection of the recoil is unaffected by it, and the probability of detecting the recoil does not depend on $X'$.   A situation with some similarities was noted by Smirnov and Zatsepin \cite{Smirnov:1991eg} for neutrinos emitted in $Z^0$ decay, where both particles are neutrinos.

\section{Conclusions}
 
We turn to a summary of our conclusions. Taken in isolation, a propagating neutrino state in a superposition of mass eigenstates may be assigned a well-defined energy or a well-defined momentum, but not both.  Equally troubling, the appropriate way to make Lorentz transformations depends on the choice.  Ignoring those difficulties, it is possible to derive in a simple way the standard form for neutrino oscillations (see \cite{Dolgov:1997xr} for a summary of such derivations).      On the other hand, when the neutrino and its associated recoil are treated as a {\em single state} whose components are entangled the above objections drop away.    Oscillations emerge as entirely a kinematic effect  observable as a spatial pattern in the wavefunction of the two-body entangled state, and Lorentz transformations are unambiguous.   

In a previous version of this paper, we found the expression for the wavelength of oscillations in  the parent-recoil separation coordinate and erroneously applied that to the oscillation probability of the neutrino detected at some distance from a source whose location was approximately known.  The wavelength in the neutrino-recoil-separation coordinate differs from the standard form of the wavelength in the neutrino-origin coordinate for neutrino oscillations.   In fact, while the two wavelengths {\em are} different, they are related in a way that makes them completely consistent \cite{Vogel,Kayser}.  Entanglement in this case therefore produces no readily observable physical consequences.    In principle, a measurement of the recoil momentum would permit an unambiguous prediction of the eigenmass detected in a neutrino detector, but such precision has not yet been realized, nor are neutrino coincidence experiments simple to do.  It is often objected that entanglement does not apply to situations where a `measurement' is made `before' the neutrino is detected, because the recoil hits something.  We would argue that the central requirement in a valid quantum-mechanical analysis is to deal with an isolated system, and it is that which forces entanglement on us in the first place.  It may be formally necessary to enlarge the system to include subsequent interactions of the recoil.  The conservation of energy that removes the energy from the phase still applies.  The momentum relationship between neutrino components will apply only under the condition that the surroundings play no role until the neutrino and recoil are beyond the range of their interaction.  There are clearly cases where that is not a good assumption, for example, the known influence of a solid crystalline medium on beta decay \cite{Arnaboldi:2006zz}.  Moreover, a complex system that encompasses the recoil and has internal degrees of freedom would technically invalidate the assumption that   $\left |R\right \rangle$ has no dependence on the index $i$, although the range argument could be invoked to rescue it.  It is therefore fortunate, perhaps, that entanglement makes no alteration to the standard form of neutrino oscillations. 

The momenta and energies of components of a propagating neutrino state have a defined relationship that is fixed by the properties of the associated recoil particle.  This relationship leads to a group velocity that corresponds to the velocity of the neutrino.  The interference of the components emerges even in a plane-wave description and is due to the differences in momenta.  The oscillation probability has a spatially periodic dependence on the separation of the neutrino and its associated recoil.  The wavelength depends on the masses of the parent and recoil, and on the Lorentz boost of the parent.   However, transformed into coordinates for the position of the neutrino relative to the decay point, the oscillation probability corresponds to the standard form and depends on neither the masses of the parent and recoil nor  on the Lorentz boost.

  \begin{acknowledgments}
The author is very grateful to the many individuals who have provided their insight, including E. Akhmedov, N. Bell, L. Bodine,  D. Boyanovsky, Y. Declais, M. Diwan, T. Goldman, M. Goodman, R. Holman, J. Hutasoit, B. Kayser, B. Keister, J. Kopp, J. Lowe, G. Miller, A. Nelson, S. Oser, A. Smirnov,  P. Vogel,  D. Wark, and J. Wu.  Any shortcomings in our work are entirely our own responsibility.    This research was supported by DOE under grant DE-FG02-97ER41020.
 \end{acknowledgments}
 
\bibliography{testbib}{}

\end{document}